\documentclass[twocolumn,showpacs,preprintnumbers,amsmath,amssymb,floatfix,superscriptaddress]{revtex4}
\newcommand{ \be }{\begin{equation}}     
\newcommand{ \ee }{\end{equation}}     
\newcommand{ \bea }{\begin{eqnarray}}     
\newcommand{ \eea }{\end{eqnarray}}

\usepackage{graphicx}% Include figure files
\usepackage{dcolumn}% Align table columns on decimal point
\usepackage{bm}% bold math

\begin{document}

\title{Incident Energy Dependence of $p_{t}$ Correlations at RHIC}

\affiliation{Argonne National Laboratory, Argonne, Illinois 60439}
\affiliation{University of Bern, 3012 Bern, Switzerland}
\affiliation{University of Birmingham, Birmingham, United Kingdom}
\affiliation{Brookhaven National Laboratory, Upton, New York 11973}
\affiliation{California Institute of Technology, Pasadena, California 91125}
\affiliation{University of California, Berkeley, California 94720}
\affiliation{University of California, Davis, California 95616}
\affiliation{University of California, Los Angeles, California 90095}
\affiliation{Carnegie Mellon University, Pittsburgh, Pennsylvania 15213}
\affiliation{Creighton University, Omaha, Nebraska 68178}
\affiliation{Nuclear Physics Institute AS CR, 250 68 \v{R}e\v{z}/Prague, Czech Republic}
\affiliation{Laboratory for High Energy (JINR), Dubna, Russia}
\affiliation{Particle Physics Laboratory (JINR), Dubna, Russia}
\affiliation{University of Frankfurt, Frankfurt, Germany}
\affiliation{Institute of Physics, Bhubaneswar 751005, India}
\affiliation{Indian Institute of Technology, Mumbai, India}
\affiliation{Indiana University, Bloomington, Indiana 47408}
\affiliation{Institut de Recherches Subatomiques, Strasbourg, France}
\affiliation{University of Jammu, Jammu 180001, India}
\affiliation{Kent State University, Kent, Ohio 44242}
\affiliation{Lawrence Berkeley National Laboratory, Berkeley, California 94720}
\affiliation{Massachusetts Institute of Technology, Cambridge, MA 02139-4307}
\affiliation{Max-Planck-Institut f\"ur Physik, Munich, Germany}
\affiliation{Michigan State University, East Lansing, Michigan 48824}
\affiliation{Moscow Engineering Physics Institute, Moscow Russia}
\affiliation{City College of New York, New York City, New York 10031}
\affiliation{NIKHEF and Utrecht University, Amsterdam, The Netherlands}
\affiliation{Ohio State University, Columbus, Ohio 43210}
\affiliation{Panjab University, Chandigarh 160014, India}
\affiliation{Pennsylvania State University, University Park, Pennsylvania 16802}
\affiliation{Institute of High Energy Physics, Protvino, Russia}
\affiliation{Purdue University, West Lafayette, Indiana 47907}
\affiliation{University of Rajasthan, Jaipur 302004, India}
\affiliation{Rice University, Houston, Texas 77251}
\affiliation{Universidade de Sao Paulo, Sao Paulo, Brazil}
\affiliation{University of Science \& Technology of China, Anhui 230027, China}
\affiliation{Shanghai Institute of Applied Physics, Shanghai 201800, China}
\affiliation{SUBATECH, Nantes, France}
\affiliation{Texas A\&M University, College Station, Texas 77843}
\affiliation{University of Texas, Austin, Texas 78712}
\affiliation{Tsinghua University, Beijing 100084, China}
\affiliation{Valparaiso University, Valparaiso, Indiana 46383}
\affiliation{Variable Energy Cyclotron Centre, Kolkata 700064, India}
\affiliation{Warsaw University of Technology, Warsaw, Poland}
\affiliation{University of Washington, Seattle, Washington 98195}
\affiliation{Wayne State University, Detroit, Michigan 48201}
\affiliation{Institute of Particle Physics, CCNU (HZNU), Wuhan 430079, China}
\affiliation{Yale University, New Haven, Connecticut 06520}
\affiliation{University of Zagreb, Zagreb, HR-10002, Croatia}

\author{J.~Adams}\affiliation{University of Birmingham, Birmingham, United Kingdom}
\author{M.M.~Aggarwal}\affiliation{Panjab University, Chandigarh 160014, India}
\author{Z.~Ahammed}\affiliation{Variable Energy Cyclotron Centre, Kolkata 700064, India}
\author{J.~Amonett}\affiliation{Kent State University, Kent, Ohio 44242}
\author{B.D.~Anderson}\affiliation{Kent State University, Kent, Ohio 44242}
\author{D.~Arkhipkin}\affiliation{Particle Physics Laboratory (JINR), Dubna, Russia}
\author{G.S.~Averichev}\affiliation{Laboratory for High Energy (JINR), Dubna, Russia}
\author{S.K.~Badyal}\affiliation{University of Jammu, Jammu 180001, India}
\author{Y.~Bai}\affiliation{NIKHEF and Utrecht University, Amsterdam, The Netherlands}
\author{J.~Balewski}\affiliation{Indiana University, Bloomington, Indiana 47408}
% ...

\collaboration{STAR Collaboration}\noaffiliation

\begin{abstract}
We present results for
two-particle transverse momentum correlations,
$\langle \Delta p_{t,i} \Delta p_{t,j} \rangle$,
as a function of event centrality for Au+Au collisions 
at $\sqrt{s_{NN}}$ = 20, 62, 130, and 200 GeV at the Relativistic 
Heavy Ion Collider.
We observe 
correlations 
decreasing with centrality that are similar at all four incident energies.  
The correlations multiplied by the multiplicity density increase 
with incident energy and the centrality dependence may show evidence 
of processes such as thermalization, jet production, or the saturation of transverse flow.  
The square root of the correlations divided by the event-wise average transverse momentum per event
shows little or no beam energy dependence and generally agrees with 
previous measurements at the Super Proton Synchrotron. 
\end{abstract}

\maketitle

The study of event-by-event 
fluctuations in global quantities, which are intimately related 
to correlations in particle production,
may provide evidence for 
the production of the quark gluon plasma (QGP) in relativistic heavy 
ion collisions \cite{stephanov_fluc_tricritical, 
stephanov_fluc_qcd_crit, fluc_collective, signatures, charge_fluct, 
charge_fluct2, balance_theory, fluctuations_review_heiselberg, fluct3, 
fluct4, fluct5, methods_fluctuations, stephanov_thermal_fluc_pion, 
hijing_jet_study, gavin_pt_fluc}.  
Various theoretical work predicts that the production of a QGP phase 
in relativistic heavy ion collisions could produce significant 
dynamic
event-by-event 
fluctuations in 
apparent temperature, mean
transverse momentum, multiplicity, and conserved 
quantities such as net charge. 
Several recent experimental studies at the SPS \cite{ceres_pt, 
wa98_fluc, na49_fluc} and at RHIC \cite{star_pt_fluc,
star_charge_fluc, 
star_balance, phenix_net_charge_fluc, phenix_pt_fluc,phenix_pt_2004} have focused 
on the study of 
fluctuations and
correlations in relativistic heavy ion collisions.  
One possible signal of the QGP would be a non-monotonic change in $p_t$ 
correlations as function of centrality and/or as the incident energy is raised \cite{fluctuations_review_heiselberg}.

Here we report an experimental  study of the incident energy dependence of $p_{t}$ correlations using Au+Au collisions ranging in center of mass energy from the highest SPS energy to the highest RHIC energy using the STAR detector at RHIC.

Fluctuations involve a purely statistical component arising from the 
stochastic nature of particle production and detection processes, 
as well as a dynamic component determined by correlations arising 
in various particle production processes.  
In this paper we first unambiguously demonstrate the existence of 
a finite dynamical component at all four incident energies by
comparing the distribution of measured event-wise average transverse momentum per
event, $\langle p_{t} \rangle$, with the same quantity from mixed
events.  
We then analyze these dynamical
fluctuations using the two particle 
transverse momentum
correlations defined as covariance
\begin{eqnarray}
\langle \Delta p_{t,i} \Delta p_{t,j} \rangle=\frac{1}{N_{\rm event}}
\sum_{k=1}^{N_{\rm event}}{\frac{C_{k}}{N_{k}(N_{k}-1)}}
\end{eqnarray}
where
\begin{eqnarray}
C_{k}=\sum\limits_{i=1}^{N_{k}} {\sum\limits_{j=1,i\ne j}^{N_{k}} 
{{\left( {p_{t,i}-\left\langle {\left\langle {p_t} \right\rangle } 
\right\rangle } \right)\left( {p_{t,j}-\left\langle 
{\left\langle {p_t} \right\rangle } \right\rangle } \right)} }}
\label{eq:three}
\end{eqnarray}
%
%\noindent
and $N_{\rm event}$ is the number of events, $p_{t,i}$ is the transverse 
momentum of the $i^{th}$ track in each event, $N_k$ is the number of 
tracks in the $k^{th}$ event.  
The 
overall event 
average transverse momentum $\left\langle {\left\langle {p_t}
\right\rangle } \right\rangle$ is given by
\begin{eqnarray}
\left\langle {\left\langle {p_t} \right\rangle } \right\rangle =\left( {\sum\limits_{k=1}^{N_{\rm event}} {\left\langle {p_t} \right\rangle _{k}}} \right)/N_{\rm event}
\label{eq:one}
\end{eqnarray}
%
%\noindent
where $\left\langle {p_t} \right\rangle _k$ is the average 
transverse momentum per event given by
\begin{eqnarray}
\left\langle {p_t} \right\rangle _k=\left(\sum\limits_{i=1}^{N_{k}} 
{p_{t,i}}\right) /N_k.
\label{eq:two}
\end{eqnarray}

\begin{figure}
\includegraphics[width=3in]{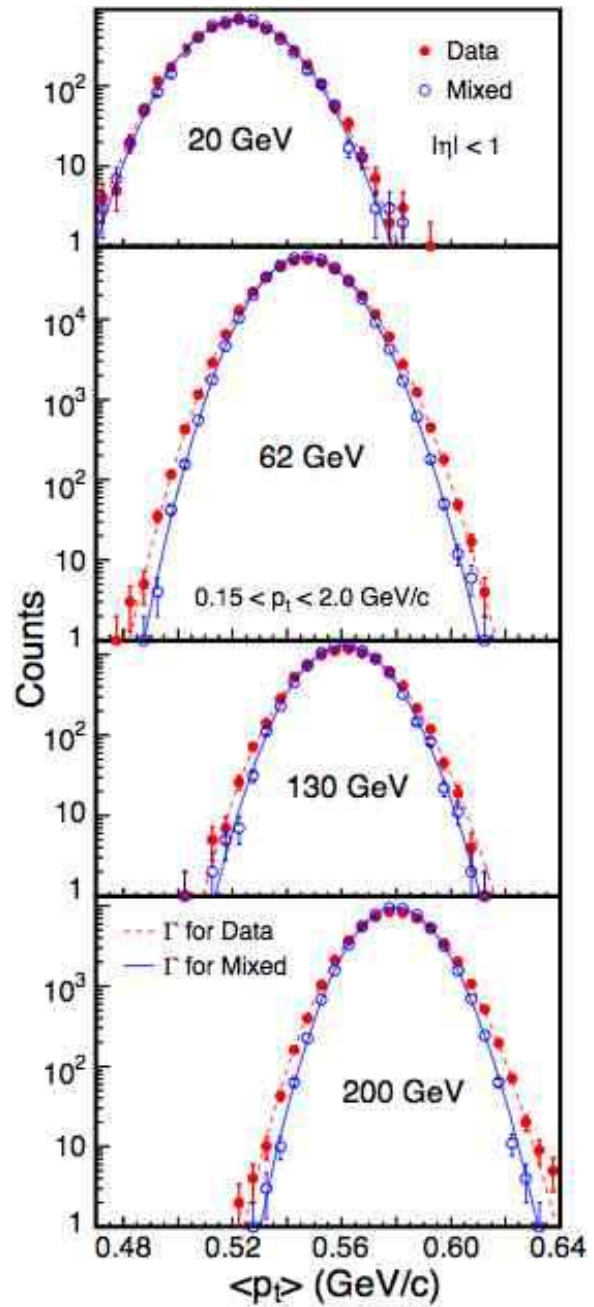}
\caption{\label{fig:fig1} (Color online) Histograms of the average transverse
momentum per event for Au+Au at $\sqrt{s_{NN}}$ = 20, 62, 130, 
and 200 GeV for the 5\% most central collisions at each energy.  
Both data and mixed events are shown for each incident energy.  
The lines represent gamma distributions.}
\end{figure}

$\langle \Delta p_{t,i} \Delta p_{t,j} \rangle$ 
is independent, to first order, of detection efficiencies
because both the numerator $C_{k}$ and 
the denominator $N_{k}(N_{k}-1)$ are proportional to the the square 
of the particle detection efficiency.  Therefore the efficiency cancels.
By construction $\langle \Delta p_{t,i} \Delta p_{t,j} \rangle$
is zero within statistics for properly mixed events because all correlations are removed.
Note that we use mixed events only in Figure 1.

The data used in this analysis were measured using the Solenoidal 
Tracker at RHIC (STAR) detector to study Au+Au collisions 
at $\sqrt{s_{NN}}$ = 20, 62, 130, and~200 GeV \cite{TPCref}.  
The main detector was the Time Projection Chamber (TPC) located 
in a solenoidal magnetic field. The magnetic field was 0.25~T for 
the 20 and 130~GeV data and 0.5~T for the 62 and 200~GeV data.  
Tracks from the TPC with 0.15~GeV/c $\le p_{t} \le$ 2.0 GeV/c 
with $|\eta| < 1.0$ were used in the analysis.  
All tracks were required to have originated within 1 cm of the 
measured event vertex.  Events were selected according to their 
distance of the event vertex from the center of STAR.  
Events were accepted within 1 cm of the center of STAR in the plane
perpendicular to the beam direction.  
For the 20 and 130~GeV data sets, events were accepted with vertices 
within 75~cm of the center of STAR in the beam direction, 
while for the 62 and 200~GeV data sets, events were accepted 
within 25~cm of the center.

Data shown for 62, 130 and 200~GeV are from minimum bias triggers.  
Minimum bias triggers were defined by the coincidence of two Zero 
Degree Calorimeters (ZDCs) \cite{ZDC} located $\pm$ 18~m from 
the center of the interaction region.  
For 20~GeV a combination of minimum bias and central triggers was used.  
Centrality bins were determined using the multiplicity of all 
charged particles measured in the TPC with $|\eta| < 0.5$.  
The centrality bins were calculated as fractions of this multiplicity 
distribution starting with the highest multiplicities.  
The ranges used were 0-5\% (most central), 5-10\%, 10-20\%, 
20-30\%, 30-40\%, 40-50\%, 50-60\%, 60-70\%, 
and 70-80\% (most peripheral).  
Each centrality was associated with a number of participating 
nucleons, $N_{part}$, using a Glauber Monte Carlo calculation \cite{npart_ref}.

We treated the variation of $\left\langle {\left\langle {p_t} \right\rangle } \right\rangle$
within a given centrality bin using the following procedure.  We calculated $\left\langle {\left\langle {p_t} \right\rangle } \right\rangle$ as a function of $N_{ch}$, the multiplicity used to define the centrality bin.  We fitted this dependence and used the fit in Eqs. 1-4 on an event-by-event basis as a function of $N_{ch}$.
This method removes the dependence of the 
experimental results on the size of the centrality bin and slightly 
reduces $\langle \Delta p_{t,i} \Delta p_{t,j} \rangle$ by removing correlations 
induced by the changing of $\left\langle {\left\langle {p_t} 
\right\rangle } \right\rangle$ within the experimental centrality bins.  
The results presented in the paper are obtained using this fitting procedure.

\begin{figure}
\includegraphics[width=3in]{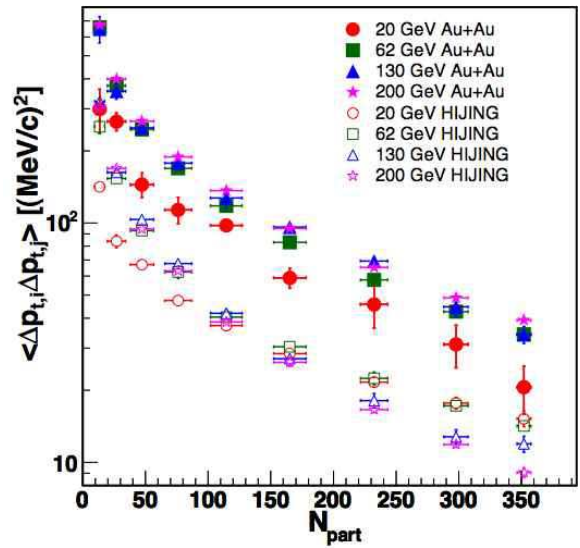}
\caption{\label{fig:fig2}(Color online) $\langle \Delta p_{t,i} \Delta p_{t,j}
\rangle$ as a function of centrality and incident energy for Au+Au 
collisions compared with HIJING results.}
\end{figure}

Fig.~\ref{fig:fig1} shows histograms of  
$\langle p_{t} \rangle$  for the 5\% most central Au+Au collisions 
at 20, 62, 130, and 200~GeV.  
Histograms for $\langle p_{t} \rangle$ are also shown for mixed
events.  The histograms for the data are wider than the histograms 
for mixed events indicating that we observe non-statistical 
fluctuations at all four incident energies.  
Similar results are obtained for all centralities.  
The overall normalization reflects the number events taken at 
each energy.  The values of $p_{t}$ included in these histograms 
are not corrected for experimental momentum resolution, acceptance or efficiency.

The mixed events at each energy were created by randomly selecting one track from 
an event chosen from measured events in the same centrality and event vertex bin.  Ten centrality bins and either five or ten bins (depending on the available number of events at each energy) in the event vertex position in the beam direction were used to create mixed events with
the same multiplicity distribution as the real events.  Note that we do not use mixed
events for the quantitative analysis based on $\langle \Delta p_{t,i} \Delta p_{t,j} \rangle$.

The lines in Fig.~\ref{fig:fig1} represent gamma distributions 
for both the data and mixed events.  The parameters for the gamma 
distributions are shown in Table 1.
According to 
Ref.~\cite{tannenbaum}, without $p_{t}$ cuts the parameter $\alpha$ divided by 
the average multiplicity in the centrality bin, 
$\langle N \rangle$, should be approximately two and 
the parameter $\beta$ multiplied by $\langle N \rangle$ should 
reflect the temperature parameter of the $p_{t}$ distributions.  
We find that $\alpha/\langle N \rangle$ varies from 2.27 to 1.93 
and $\beta\langle N \rangle$ varies from 0.230 to 0.299 GeV/c 
as the energy goes from 20 to 200 GeV.
 
\begin{table}[htdp]
\caption{Parameters for the gamma distributions shown in Fig. 1.  
The gamma distribution is given by the form 
$f(x)=\frac{x^{\alpha -1}e^{-x/\beta }} {\Gamma \left( \alpha  
\right)\beta ^\alpha }$ where $\alpha =\frac{\mu ^2} {\sigma ^2}$ 
and $\beta =\frac{\sigma ^2}{\mu}$ in GeV/c.  
$\mu$ is the mean in GeV/c and $\sigma$ is the standard deviation in GeV/c.}
\begin{center}
\begin{tabular}{|l|c|c|c|c|} \hline
Case			& $\alpha$ 	& $\beta$ 				& $\mu$ 		& $\sigma$ \\ \hline
20 GeV Real		& 1096 		& 4.772 x 10$^{-4}$ 		& 0.5228 		& 0.01579 \\
20 GeV Mixed		& 1199 		& 4.360 x 10$^{-4}$ 		& 0.5227 		& 0.01510 \\
62 GeV Real		& 1445 		& 3.786 x 10$^{-4}$ 		& 0.5471 		& 0.01439 \\
62 GeV Mixed		& 1743 		& 3.139 x 10$^{-4}$ 		& 0.5470 		& 0.01310 \\
130 GeV Real		& 1556 		& 3.608 x 10$^{-4}$ 		& 0.5614 		& 0.01423 \\
130 GeV Mixed		& 1917 		& 2.927 x 10$^{-4}$ 		& 0.5612 		& 0.01282 \\
200 GeV Real		& 1853 		& 3.129 x 10$^{-4}$ 		& 0.5799 		& 0.01347 \\
200 GeV Mixed		& 2373 		& 2.443 x 10$^{-4}$ 		& 0.5799 		& 0.01190 \\ \hline
\end{tabular}
\end{center}
\label{default}
\end{table}%

To characterize the 
transverse momentum
correlations, we use the the quantity 
$\langle \Delta p_{t,i} \Delta p_{t,j} \rangle$, defined in Eq. 1. 
Fig.~\ref{fig:fig2} shows 
$\langle \Delta p_{t,i} \Delta p_{t,j} \rangle$ for Au+Au collisions 
at $\sqrt{s_{NN}}$ = 20, 62, 130, and 200 GeV as a function of
centrality.  
One observes that $\langle \Delta p_{t,i} \Delta p_{t,j} \rangle$ 
decreases with centrality at all four energies as expected 
due to
a progressive dilution of the correlations resulting from 
the increased number of participants if
the correlations are dominated by pairs of particles
that originate from the same nucleon-nucleon collision.
The correlations 
measured at 62, 130 and 200 GeV are similar 
while the correlations  
for 20 GeV are smaller than those observed at the higher energies.

\begin{figure}
\includegraphics[width=3in]{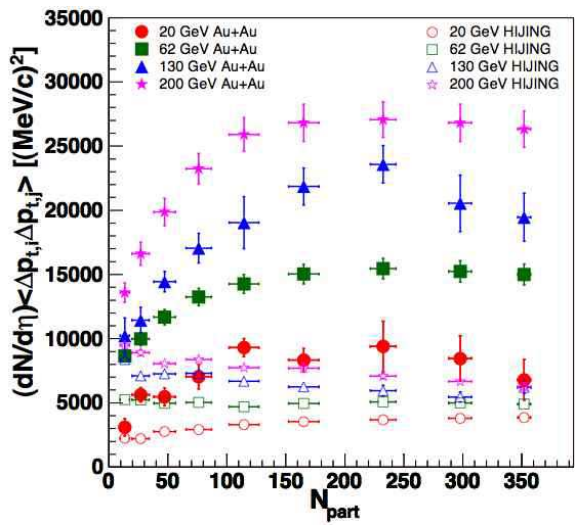}
\caption{\label{fig:fig3}(Color online) $(dN/d\eta)\langle \Delta p_{t,i} 
\Delta p_{t,j} \rangle$ as a function of centrality and incident 
energy for Au+Au collisions compared with HIJING results.}
\end{figure}

\begin{figure}
\includegraphics[width=3in]{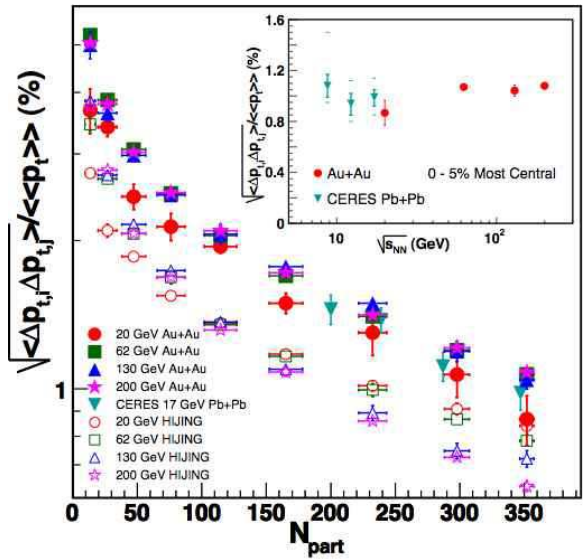}
\caption{\label{fig:fig4}(Color online) $\sqrt{\langle \Delta p_{t,i} 
\Delta p_{t,j} \rangle}/\left\langle {\left\langle {p_t} 
\right\rangle } \right\rangle$ as a function of centrality 
and incident energy for Au+Au collisions compared with 
HIJING results for corresponding systems.  
The inset shows the excitation function for the most central bin.}
\end{figure}

To explore the issue of the relative importance of short range
correlations such as Coulomb 
interactions
and Hanbury Brown-Twiss (HBT) effects, we extracted the correlations excluding pairs with 
invariant relative momentum, $q_{inv}$, less than 0.1~GeV/c, 
assuming that all particles were pions.  
We observed that 10\% of the measured correlations at 62, 130, and
200~GeV and 20\% of measured correlations at 20~GeV could 
be attributed to these short range correlations.  
These estimates agree with those extracted for 
17~GeV Pb+Pb~\cite{ceres_pt} using a 
somewhat different method. 
We also estimated  
the contribution of resonances and other charge-ordering
effects by studying the reduction in 
the correlations for 
same charge (negative) 
particles compared with correlations for all charged particles.  This study indicated that the
reduction in $\langle \Delta p_{t,i} \Delta p_{t,j} \rangle$ is 40\% at 20~GeV, 20\% at 62 and 130~GeV, and 15\%  at 200~GeV.
We do not correct $\langle \Delta p_{t,i} \Delta p_{t,j} \rangle$
for short range correlations or resonance contributions.

The errors shown 
in all figures
are statistical unless otherwise noted.  
We estimate the 
systematic relative errors for $\langle \Delta p_{t,i} \Delta p_{t,j} \rangle$ 
using studies of the effects of $p_{t}$-dependent efficiencies
(1.2\%) and
sensitivity to track merging and splitting (1.4\%).
These values give an 
overall
systematic relative error 
of 2\%.
The measured correlations were lowered approximately 3\% using the 
fitting method rather than the binning method.  
The reported values are sensitive to the $p_{t}$ cuts for kinematic and physics reasons.  Using HIJING \cite{hijing} we observe a 6\% increase in correlations when the lower $p_{t}$ cut is removed.  Raising the upper $p_{t}$ cut increases the correlations.  We used 0.15~GeV/c $\le p_{t} \le$ 2.0 GeV/c for all the results reported in this paper.  The upper $p_{t}$ cut was chosen to be consistent with previous work \cite{star_pt_fluc,phenix_pt_2004}.

Also shown in  Fig.~\ref{fig:fig2} are 
HIJING calculations for Au+Au collisions 
at $\sqrt{s_{NN}}$ = 20, 62, 130, and 200~GeV \cite{hijing}.
We used HIJING version
1.36 with the default options, which includes jet quenching.  The HIJING results
were obtained by selecting particles with  0.15~GeV/c $\le p_{t} \le$ 2.0 GeV/c 
with $|\eta| < 1.0$\ without further efficiency corrections.
HIJING reproduces correlations in p+p and $\alpha$+$\alpha$ collisions at Intersecting Storage Ring (ISR) energies \cite{ISR_data}, p+p collisions at RHIC energies, and p+$\bar{\rm p}$ collisions at CERN p+$\bar{\rm p}$ Collider (SppS)  energies \cite{HIJING_verification}.  We use HIJING to provide a reference incorporating a superposition of nucleon-nucleon interactions.
Any differences between HIJING and the experimental results might signal phenomena unique to nucleus-nucleus collisions.
The HIJING calculations exhibit little incident energy dependence 
and decrease with increasing centrality.  
The values for $\langle \Delta p_{t,i} \Delta p_{t,j} \rangle$
predicted by HIJING are always smaller than the data.

To address the observed 
dilution of the correlations with centrality, and to check the
hypothesis that the correlations scale as inverse multiplicity,  
we multiply$\langle \Delta p_{t,i} \Delta p_{t,j} \rangle$ by the 
charged particle pseudorapidity density at a given centrality,
$dN/d\eta$.  
We 
use fully corrected values for $dN/d\eta$ from published work 
\cite{phobos_dNdeta1,phobos_dNdeta2,phobos_dNdeta3}. 
The quantity $\frac{dN}{d\eta}\langle \Delta p_{t,i} 
\Delta p_{t,j} \rangle$ then is insensitive to efficiency 
and is similar to the 
(efficiency corrected)
quantity $\Delta \sigma _{pt}$~\cite{star_pt_fluc} 
that STAR has 
reported 
previously.

In  Fig.~\ref{fig:fig3} we show the quantity 
$\frac{dN}{d\eta}\langle \Delta p_{t,i} \Delta p_{t,j} \rangle$ 
for Au+Au collisions at 20, 62, 130, and 200 GeV as 
a function of centrality.  In this figure the errors include the quoted
errors in $dN/d\eta$.  
This quantity increases with incident energy 
at all centralities.  
At each energy this measure of the correlations increases quickly as 
the collisions become more central and then saturates in central
collisions.  
The behavior of this quantity is similar to that of 
the quantity $\Delta \sigma _{pt}$ previously studied 
by STAR \cite{star_pt_fluc}. 
This saturation might indicate effects such as the onset 
of thermalization \cite{gavin_pt_fluc}, 
the onset of jet quenching \cite{hijing_jet_study}, 
the saturation of  
transverse flow 
\cite{voloshin} in central collisions, or other processes.

In Fig.~\ref{fig:fig3} 
the results of
HIJING calculations for $\frac{dN}{d\eta}\langle \Delta p_{t,i} 
\Delta p_{t,j} \rangle$ are also shown.  
In contrast to the experimental results, the HIJING results 
show little dependence on centrality.

To account for possible changes of 
$\langle \Delta p_{t,i} \Delta p_{t,j} \rangle$ 
due to possible changes in 
 $\left\langle {\left\langle {p_t} \right\rangle } \right\rangle$
with incident energy and/or centrality of the collision, we also study
the square root of the measured 
correlations 
scaled
by $\left\langle {\left\langle {p_t} \right\rangle } \right\rangle$.  
The resulting quantity $\sqrt{\langle \Delta p_{t,i} \Delta p_{t,j} 
\rangle}/\left\langle {\left\langle {p_t} \right\rangle }
\right\rangle$ is shown in Fig.~\ref{fig:fig4} for Au+Au collisions 
at 20, 62, 130, and 200~GeV.  
Similar results from Pb+Pb collisions at 17~GeV \cite{ceres_pt} 
are also shown in Fig.~\ref{fig:fig4}.  
These values are consistent with our measured results for Au+Au 
at 20~GeV.  
We observe little or no dependence on the incident energy for this quantity.  
The inset in Fig.~\ref{fig:fig4} demonstrates the incident 
energy dependence of $\sqrt{\langle \Delta p_{t,i} \Delta p_{t,j} 
\rangle}/\left\langle {\left\langle {p_t} \right\rangle } 
\right\rangle$ for the 0 - 5\% most central bin where the Pb+Pb
results are from Ref.~\cite{ceres_pt}.  

In contrast to the measured correlations, HIJING predictions 
for $\sqrt{\langle \Delta p_{t,i} \Delta p_{t,j} \rangle}/\left\langle 
{\left\langle {p_t} \right\rangle } \right\rangle$ vary with incident energy.  HIJING predicts a different centrality dependence as well as a noticeable dependence on the incident energy.

In conclusion we observe clear  
non-zero $p_t$ correlations,
$\langle \Delta p_{t,i} \Delta p_{t,j} \rangle$, in 
 Au+Au
collisions from $\sqrt{s_{NN}}$ = 20 to 200 GeV.   
The quantity  $\frac{dN}{d\eta}\langle \Delta p_{t,i} \Delta p_{t,j} 
\rangle$ increases with beam energy.  
The centrality dependence of  $\frac{dN}{d\eta}\langle 
\Delta p_{t,i} \Delta p_{t,j} \rangle$ may show signs of
effects such as 
thermalization \cite{gavin_pt_fluc}, 
the onset of jet suppression \cite{hijing_jet_study,phenix_pt_2004},
the saturation of transverse expansion in central collisions \cite{voloshin},
or other processes.  
The quantity  
$\sqrt{\langle \Delta p_{t,i} \Delta p_{t,j} 
\rangle}/\left\langle {\left\langle {p_t} \right\rangle } 
\right\rangle$ shows little or no change with beam energy.  
HIJING 
model calculations
underpredict the measured correlations and 
do not predict the observed centrality dependence.

We thank the RHIC Operations Group and RCF at BNL, and the
NERSC Center at LBNL for their support. This work was supported
in part by the HENP Divisions of the Office of Science of the U.S.
DOE; the U.S. NSF; the BMBF of Germany; IN2P3, RA, RPL, and
EMN of France; EPSRC of the United Kingdom; FAPESP of Brazil;
the Russian Ministry of Science and Technology; the Ministry of
Education and the NNSFC of China; IRP and GA of the Czech Republic,
FOM of the Netherlands, DAE, DST, and CSIR of the Government
of India; Swiss NSF; the Polish State Committee for Scientific 
Research; and the STAA of Slovakia.

\end{document}